\documentclass{emulateapj}
\input{epsf}
\submitted{Accepted for publication in ApJ Letters on April 19, 2012}
\shorttitle{
How to make an ultra-faint dwarf galaxy}

\shortauthors{E. L. {\L}okas et al.}

\begin{document}

\title{How to make an ultra-faint dwarf spheroidal galaxy:\\ tidal stirring of
  disky dwarfs with shallow dark matter density profiles}

\author{Ewa L. {\L}okas\altaffilmark{1}, Stelios
  Kazantzidis\altaffilmark{2} and Lucio Mayer\altaffilmark{3}}

\altaffiltext{1}{Nicolaus Copernicus Astronomical Center, 00-716 Warsaw, Poland; lokas@camk.edu.pl}
\altaffiltext{2}{Center for Cosmology and Astro-Particle Physics; and Department of Physics; and Department of Astronomy,
    The Ohio State University, Columbus, OH 43210, USA; stelios@astronomy.ohio-state.edu}
\altaffiltext{3}{Institute for Theoretical Physics, University of Z\"urich, CH-8057 Z\"urich, Switzerland;
        lucio@phys.ethz.ch}

\begin{abstract}

In recent years the Sloan Digital Sky Survey has unraveled a new
population of ultra-faint dwarf galaxies (UFDs) in the vicinity of the
Milky Way (MW) whose origin remains a puzzle.
Using a suite of collisionless $N$-body simulations,
we investigate the formation of UFDs in the context of the tidal
stirring model for the formation of dwarf spheroidal galaxies in the
Local Group (LG). Our simulations are designed to reproduce the tidal
interactions between MW-sized host galaxies and rotationally supported
dwarfs embedded in $10^{9} M_{\odot}$ dark matter (DM) halos.  We
explore a variety of inner density slopes $\rho \propto
r^{-\alpha}$ for the dwarf DM halos, ranging from core-like ($\alpha =
0.2$) to cuspy ($\alpha = 1$), and different dwarf orbital
configurations. Our experiments demonstrate that UFDs can be produced
via tidal stirring of disky dwarfs on relatively tight orbits,
consistent with a redshift of accretion by the host galaxy of $z \sim
1$, and with intermediate values for the halo inner density slopes
($\rho \propto r^{-0.6}$). The inferred slopes are in excellent
agreement with those resulting from both the modeling of the rotation
curves of dwarf galaxies and recent cosmological simulations of dwarf
galaxy formation. Comparing the properties of observed UFDs with those
of their simulated counterparts, we find remarkable similarities in
terms of basic observational parameters. We conclude that tidal
stirring of rotationally supported dwarfs represents a viable
mechanism for the formation of UFDs in the LG environment.

\end{abstract}


\section{Introduction}

The Sloan Digital Sky Survey (SDSS) has revealed
a new population of dwarf galaxies in the Local Group (LG), known as
ultra-faint dwarfs (UFDs) (Willman et al. 2005a,b; Belokurov et
al. 2006, 2007, 2008, 2009, 2010; Zucker et al. 2006a,b; Walsh et
al. 2007; Irwin et al. 2007). UFDs are typically fainter than $M_V=-8$
mag and extremely metal-poor, and they are located surprisingly close
to the Milky Way
(MW), mostly within its virial radius. These systems are also
believed to be strongly dark matter (DM) dominated, even more than
the classic
MW dwarf spheroidal galaxies (dSphs) (Simon \& Geha 2007; Strigari et al. 2008).

The origin of UFDs is a matter of ongoing debate. Here, we
investigate a scenario for their formation based on the tidal
stirring model for the origin of classic dSphs in the LG (Mayer et
al. 2001). This model
relies on the gravitational interaction between initially disky
dwarfs and massive host galaxies and its ability to produce
classic dSphs
has been recently demonstrated (Kazantzidis et al. 2011; {\L}okas et
al. 2011a). In most previous numerical studies, the DM halos of the
progenitor
dwarfs followed the Navarro, Frenk \& White (1997, NFW)
density profile and none of the simulations produced an UFD in the end.
Recently, however, hydrodynamical simulations of the formation of isolated
dwarf galaxies in the cosmological context have been performed by
Governato et al. (2010). These authors modeled
inhomogeneous interstellar medium and found that the resulting gas-rich
disky dwarfs acquire a significantly shallower inner DM slope relative
to the cuspy NFW profiles expected in pure DM simulations, in
close agreement with the slopes of observed dwarfs (Oh et al. 2011).

Motivated by the aforementioned developments, we perform for the first
time a series of
tidal stirring simulations of disky dwarfs embedded in DM halos with
different inner density slopes. Our experiments show
that the properties of the
resulting systems depend very sensitively on the initial inner
slope of the DM density profile. More
specifically, we demonstrate that tidal stirring of disky dwarfs is
capable of producing stellar systems with properties akin to those of
UFDs in the LG, provided that the progenitors are placed on relatively
tight, eccentric orbits inside MW-sized hosts and are embedded in DM
halos with mild central density cusps.

\section{The simulations}

We employ the method of Widrow et al. (2008) to construct fully
self-consistent, equilibrium models of dwarf galaxies consisting of
exponential stellar disks embedded in spherical DM halos.  The DM density profiles have the form
\begin{equation}    \label{densityprofiles}
    \rho(r) = \frac{\rho_{\rm char}}{(r/r_{\rm
    s})^\alpha \,(1+r/r_{\rm s})^{3-\alpha}}
\end{equation}
with an asymptotic inner $r^{-\alpha}$
and an outer slope $r^{-3}$
({\L}okas 2002; {\L}okas \& Mamon 2003). We construct three
dwarf galaxies with the same virial mass of $M_{\rm vir} = 10^{9} M_{\odot}$
and concentration parameter $c=r_{\rm vir}/r_{\rm s}=20$, but different $\alpha$,
which requires slightly different values of the characteristic density
$\rho_{\rm char}$ in equation (\ref{densityprofiles}).
Specifically, we choose $\alpha=1$ (which corresponds to the
NFW profile), and two shallower inner slopes, namely a mild cusp with
$\alpha=0.6$ and a core-like profile with $\alpha=0.2$. We note that
these shallow slopes bracket the range of possible DM slopes inferred
from recent cosmological simulations of dwarf galaxy formation at
$1<z<2$ (Governato et al. 2010).

The DM halos were populated with stellar disks of mass given
as a fraction, $m_d$, of $M_{\rm vir}$, and we chose $m_d = 0.02$. The adopted disk radial scale
length was equal to $R_d = 0.41$ kpc corresponding to a dimensionless
spin parameter of $\lambda=0.04$ (Mo et al. 1998), and the disk
thickness was equal to $z_d/R_d = 0.2$,
where $z_d$ denotes the vertical disk scale height (see Kazantzidis et
al. 2011 regarding the choices for the values of these
parameters). For simplicity,
we also assumed a constant value for the central radial velocity
dispersion $\sigma_{R0}=10$ km s$^{-1}$ for all dwarf models.
The gross properties of the three dwarfs, including
their {\it total} masses, were very similar. The resulting disks
differed only in their planar velocity dispersions and the
Toomre $Q$ parameter. $Q$ was equal to 3.8, 3.3 and 2.9 for $\alpha=1$,
0.6 and 0.2, respectively, indicating that all disks were stable against bar formation in
isolation.

Figure~1 illustrates different properties
of our progenitor dwarf models. Although our modeling approach is not unique, our
choice of density slopes ensures that the employed DM density
profiles (upper panel) have significantly different shapes in the
region occupied by the bulk of the stellar component ($\lesssim
2$~kpc), a fact which will be crucial for the interpretation of the
results. Moreover, our choices result in dwarf galaxies with
different maximum circular velocities, $V_{\rm max}$ (middle panel),
with the steeper $\alpha$ corresponding to a larger $V_{\rm max}$
($V_{\rm max} = 19.6$, 17.8 and 16.7 km s$^{-1}$, respectively, for
$\alpha=1$, 0.6 and 0.2). This is important as smaller $V_{\rm max}$
are associated with lower densities in the inner regions and,
correspondingly, longer internal dynamical times. Thus, the dwarfs
with shallow inner DM slopes are expected to respond more
impulsively to the external tidal perturbation and tidal heating
will be particularly efficient (Gnedin et al. 1999). Lastly, the
binding energy of stars (lower panel) is another key factor that
determines their response to tidal shocks. As a result, we
expect the dwarfs with shallow cusp slopes to experience much stronger
tidal evolution compared to their counterparts with steep
cusp slopes.

\begin{figure}
\begin{center}
    \leavevmode
    \epsfxsize=7.2cm
    \epsfbox[0 0 160 480]{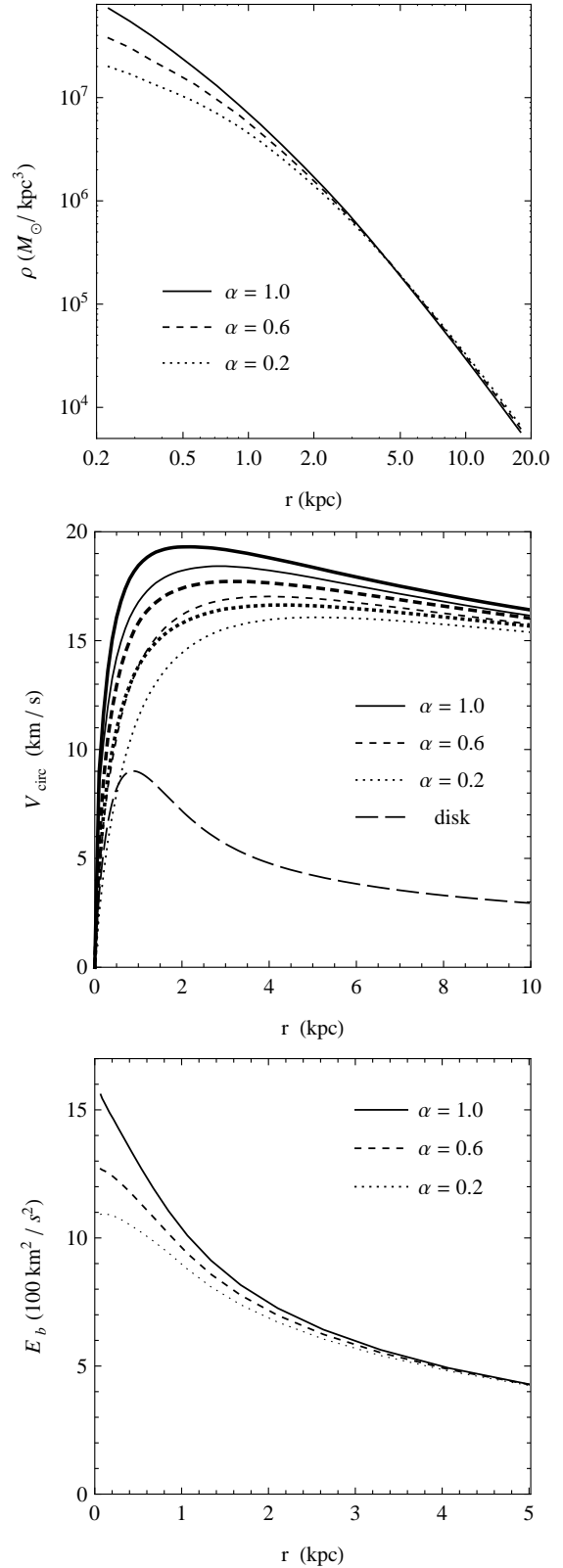}
\end{center}
\caption{The dependence of the initial properties of the adopted dwarf
  models on the inner slope $\alpha$. Upper panel: the DM density profiles,
  middle panel: the disk, halo (thin lines), and total (thick lines) circular velocity profiles,
  lower panel: the binding energy of the stars per unit mass.}
\label{alphadependence}
\end{figure}

Each dwarf galaxy model contained $N_h = 10^6$ DM and $N_d =
5 \times 10^5$ disk particles with a gravitational softening of
$\epsilon_h=60$~pc and $\epsilon_d=20$~pc, respectively.
We assumed a single primary galaxy with the present-day
structural properties of the MW (Widrow \& Dubinski 2005; Kazantzidis et al. 2011).
Each of the three dwarfs was placed on five different bound
orbits inside the primary galaxy, for a total of $15$ numerical
experiments. Specifically, we employed orbits R1-R5 of different size
and eccentricity from Kazantzidis et al. (2011), with orbital apocenters
$r_{\rm apo}$ and pericenters $r_{\rm peri}$ given in kpc in parentheses:
R1(125, 25), R2(85, 17), R3(250, 50), R4(125, 12.5), R5(125, 50).  In
all simulations, the dwarfs were initially placed at the apocenters
and the evolution was followed for $10$~Gyr.  The orientation of the
internal angular momentum of all dwarfs with respect to the orbital
angular momentum was mildly prograde and equal to $i=45^{\circ}$.
All numerical experiments were performed with the $N$-body code PKDGRAV (Stadel 2001).

\section{The formation and properties of simulated UFDs}

The tidal evolution of our dwarfs follows the general picture established
in earlier works (see Mayer et al. 2001, 2007;
Klimentowski et al. 2007, 2009; Kazantzidis et al. 2011; {\L}okas et
al. 2011b). In particular, all dwarf galaxies undergo mass loss,
their morphology changes from a disk to a triaxial and
then to a more spherical shape, and the stellar rotation is
transformed into random motions. The degree of these changes is
sensitive to the orbital configuration, namely it is proportional to
the total tidal force experienced by the dwarfs (see {\L}okas et
al. 2011c) but also depends on the structural properties of the
progenitors (Kazantzidis et al. 2011; {\L}okas et al. 2011a,b)

\begin{table}
\caption{Parameters of the simulated UFDs}
\label{parameters}
\begin{center}
\begin{tabular}{lrr}
\hline
\hline
Parameter                            &    R2      &    R4       \\
\tableline
$r_{\rm apo}$ (kpc)                  &    85      &   125       \\
$r_{\rm peri}$ (kpc)                 &    17      &   12.5      \\
$T_{\rm orb}$ (Gyr)                  &   1.3      &   1.8       \\
$N_{\rm apo}$                        &     6      &    6        \\
$T_{\rm la}$ (Gyr)                   &   6.4      &   9.3       \\
$L_V (10^3 L_{\odot})$               &   6.38     &   4.84      \\
$M_V$ (mag)                          & $-4.68$    &  $-4.38$    \\
$\mu_V$ (mag arcsec$^{-2}$)          &  28.3      &   28.7      \\
$r_{1/2}$ (kpc)                      &   0.166    &   0.137     \\
\hline
\end{tabular}
\end{center}
\end{table}

\begin{figure}
\begin{center}
    \leavevmode
    \epsfxsize=7cm
    \epsfbox[0 0 160 465]{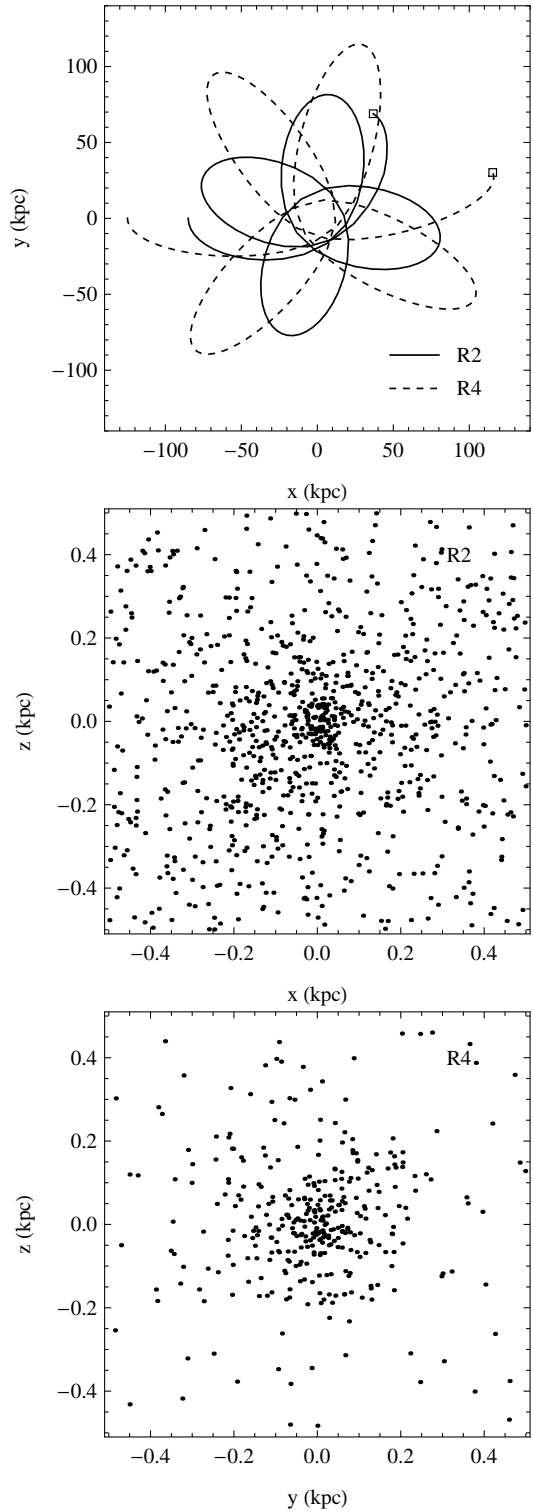}
\end{center}
\caption{Upper panel: The orbits of dwarfs in simulations R2 and R4. The orbits run anti-clockwise
  from the initial apocenter on the left up to the sixth apocenter
  marked with an open square. Middle and lower panel:
  The stars in our simulated UFD galaxies identified at the
  sixth apocenter of runs R2 and R4
  with $\alpha=0.6$. The images show 907 and 356 stars for R2 and R4
  respectively, contained in a box of 1 kpc on a side and centered on
  the dwarf. The positions of the stars were projected along the $y$
  (for R2) and $x$ (for R4) axis of the simulation box.}
\label{orbitssnapshots}
\end{figure}

Here we attempt to investigate the effect of the inner cusp
of the DM halo. We note that the dependence of the overall morphological
transformation of the dwarfs on the inner density slope will be
discussed in a future work; here we
only mention that dwarfs with shallower cusps are more susceptible to
changes, they lose mass more effectively and are thus more easily
destroyed with obvious implications for the missing satellites
problem. In particular, dwarfs with the standard NFW cusp of
$\alpha = 1.0$ survive until the end of the simulation for all orbits,
while dwarfs with the shallowest inner profile with $\alpha = 0.2$
are dissolved completely for the tighter orbits R1, R2 and R4 after 4,
3 and 2 pericenter passages,
respectively. The destruction of these
dwarfs occurs rather early in the evolution, when the dwarfs are still
quite massive, via elongation of the bar formed at an earlier
stage. On the other hand, the intermediate cusp of $\alpha = 0.6$
leads to a longer survival, but the dwarfs become much smaller in
size.

In what follows, we focus on the evolution of dwarfs with $\alpha =
0.6$ on orbits R2 and R4 which leads to the formation of dSph galaxies with
properties akin to those of present-day UFDs in the vicinity of the MW.
The two orbits
are plotted in the upper panel of Figure~2 up to the
sixth apocenter, the last one at which the dwarfs are still
discernible. The initial apo- and pericenter distances of these
orbits, $r_{\rm apo}$ and $r_{\rm peri}$, their orbital times $T_{\rm
  orb}$, number of apocenters after which the dwarfs are identified,
$N_{\rm apo}$, and times when these last apocenters took place (from
the start of the simulation), $T_{\rm la}$, are listed in the first
five rows of Table~\ref{parameters}.

The stellar components of the dwarfs at the sixth apocenter
are shown in the two lower panels of Figure~2.
The stars of the dwarfs and their
immediate surroundings were selected from a box of size 1 kpc and
their positions projected along one direction. The dwarfs thus appear
as they would be seen by a distant observer looking at them along
one axis of the simulation box if the observer was able to subtract
the contamination from the MW and the more distant tidal debris
from the dwarf itself. The very low numbers of the stars (907 and 356
for R2 and R4) mean that the dwarfs would only appear as very small
overdensities against the background (see below).

In Figure~3 we present stellar number density
profiles in the dwarfs formed in runs R2 and R4.  The density was
measured by counting stars in logarithmically spaced bins of radius
starting from 3 softening scales of stellar particles.  In each panel
the curves from top to bottom correspond to measurements performed at
subsequent apocenters, from the second to the sixth. The evolution of
the density profiles, and more specifically the decrease of the
normalization, reflects the stellar mass loss occurring as a result of
the interaction with the host galaxy.  In addition, this Figure
demonstrates that tidal shocks do not serve to modify the central
density cusp of the stellar distribution, in agreement with the study
of Kazantzidis et al. (2004) where pure DM satellites were
evolved inside a static host potential. Lastly, each density profile
shows a characteristic transition from the bound stellar component to
the tidal tails, marked by an abrupt change in the outer slope of the
profile, from the steeper to the shallower. This change occurs at
smaller radii for subsequent apocenters signifying the decreasing size
of the dwarf.

\begin{figure}
\begin{center}
    \leavevmode
    \epsfxsize=7cm
    \epsfbox[0 0 170 325]{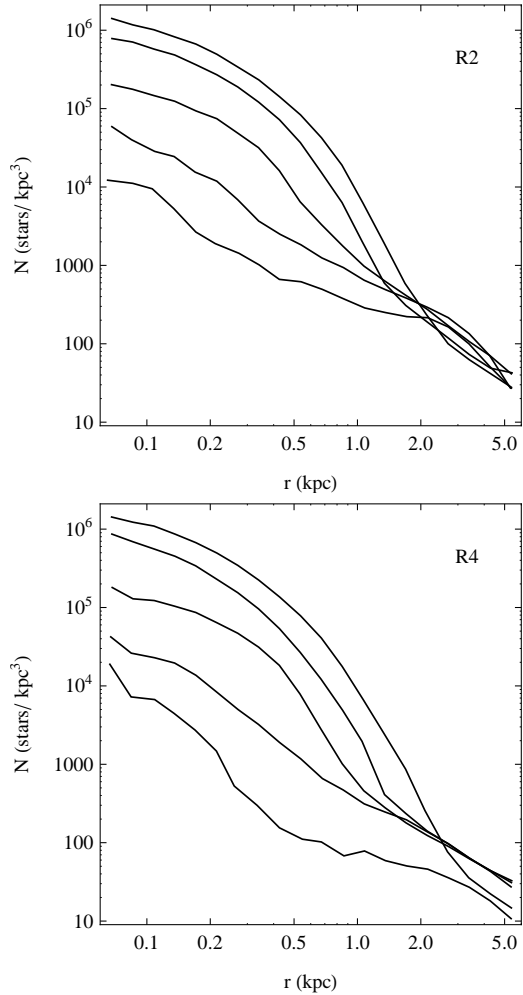}
\end{center}
\caption{Stellar number density profiles of the dwarf galaxies
  measured at subsequent apocenters for simulation R2 (upper panel)
  and R4 (lower panel) with $\alpha=0.6$. The solid lines in each panel show decreasing
  stellar densities from the second to the sixth apocenter (from the
  top to the bottom curve).}
\label{densitystars}
\end{figure}

In order to quantify this evolution we fitted each profile with
the Plummer formula estimating the half-light radius $r_{1/2}$ and
normalization. The fits were done only to data points (weighted by density)
within the range of radii unaffected by tidal tails.
The fitted normalization, which is the total
number of stars, multiplied by the stellar mass and the stellar
mass-to-light ratio, gives the total luminosity of the dwarf at
each apocenter $L_V$ or equivalently its visual magnitude $M_V$.
For the stellar mass-to-light ratio we adopt a
constant value $M/L_V = 2.5 M_\odot/L_\odot$ as predicted for the
present time for a simple low-metallicity stellar
population in the standard model described by Bruzual \& Charlot
(2003).  Thus at all stages, the luminosities of the simulated dwarfs
can be directly compared to the properties of the real MW UFDs.

We also measured the central surface brightness $\mu_V$ of the dwarfs
at each stage by first selecting the stars within $5 r_{1/2}$ from the
center and then counting stars contained within a cylinder of radius
$0.2 r_{1/2}$ along the three axes of the simulation box. The final
value follows from the average number of stars measured in the three
directions. The values of all the observational parameters of the two
dwarfs at the sixth apocenter are listed in the last four rows of
Table~\ref{parameters}.

\begin{figure}
\begin{center}
    \leavevmode
    \epsfxsize=7cm
    \epsfbox[0 0 160 320]{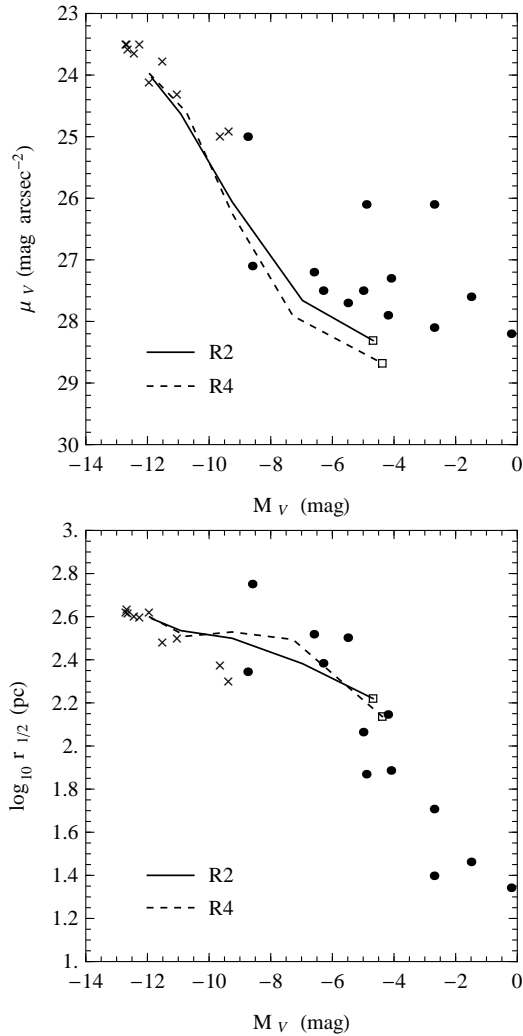}
\end{center}
\caption{Evolutionary tracks of the simulated dwarfs in runs R2 and R4
  with $\alpha=0.6$ in the $M_V$-$\mu_V$ (upper panel) and $M_V$-$r_{1/2}$ (lower panel)
  planes. The solid and dashed lines (for R2 and R4, respectively)
  join the values of the parameters measured at subsequent apocenters
  from the second to the sixth. Open squares mark the final values
  measured at the sixth apocenter. Crosses indicate observational
  parameters at the last apocenter for 10 dwarfs (on orbits R1-R5
  and with different $\alpha$) which did not end up
  as UFDs, but survived for 10 Gyr. Circles show values
  of the analogous parameters for the real UFDs of the MW.}
\label{sbradmag}
\end{figure}

We summarize the evolution of these properties by plotting in
Figure~4 the evolutionary tracks of the dwarfs in
simulations R2 and R4 in the $M_V$-$\mu_V$ and $M_V$-$r_{1/2}$
planes. We include as crosses analogous properties for 10 other
of our simulated dwarfs which survive until the end much brighter
and are thus more akin to classic dSph galaxies of the LG (see
Figure 20 in {\L}okas et al. 2011a).  We also show with circles the parameters of
the real MW UFDs, as determined by Martin et
al. (2008).  The
brightest dwarfs, with $M_V < - 8$ mag, are the Canes Venatici I dwarf
($M_V = - 8.6$) and Draco ($M_V = - 8.75$). Draco is usually
considered as one of the classic
dSphs, but its luminosity indicates that it is rather a border-line
case between the brighter population of classic dwarfs and the UFDs.

The comparison of the evolutionary tracks of the simulated dwarfs to
the real data proves that the tidal stirring indeed brings the
initially massive subhalos possessing mild inner DM cusps and
populated with a standard stellar disk to regions of parameter space
occupied by UFDs found around the MW. The end products of
runs R2 and R4, although tracing different orbits around the host, are
similar to each other and particularly close in their observational
properties to the UFD Ursa Major II with $M_V = -4.2$ mag,
$\mu_V=27.9$ mag arcsec$^{-2}$, and $r_{1/2}=0.140$
kpc (Martin et al. 2008). Although our values of the central surface
brightness are a little below the observed range, we note that the
real dwarfs may be contaminated by MW stars and therefore have their
brightnesses overestimated. Still, our simulated dwarfs should be
discernible above the background level due to the tidally stripped
stars which we estimate to be of the order of 31.2 and 32.4 mag
arcsec$^{-2}$ at $5 r_{1/2}$ for R2 and R4 respectively.

\section{Discussion}

Using collisionless $N$-body simulations, we have
demonstrated that the tidal stirring of $\sim 10^9 M_\odot$ disky
dwarfs for a period of $\sim 8$~Gyr can lead to the formation of
stellar systems with properties akin to those of UFDs around the MW.
Specifically, our experiments have shown that
UFDs are produced when the progenitor disky dwarfs are placed on
relatively tight ($r_{\rm peri} \lesssim 20$~kpc), eccentric ($r_{\rm
  apo}/r_{\rm peri} \gtrsim 5$) orbits inside a MW-sized host and are
embedded in DM halos with a mild central density cusp ($\rho \propto
r^{-0.6}$).

Such slopes agree with those inferred from both the
modeling of rotation curves of dwarf galaxies (Oh et al. 2011) and
recent cosmological simulations of dwarf galaxy formation (Governato
et al. 2010). In the latter work, cusp flattening occurs
as a result of potential fluctuations induced by rapid removal of
baryons via supernovae winds, a mechanism that depends both on the
stellar and halo mass.  Because our models lack gas, dissipation and
star formation, a direct comparison with Governato et al. is
difficult.  However, we note that the baryonic and halo masses of the
Governato et al. dwarfs at $z > 1$ are similar to those adopted in the
present study (within a factor of a few). This together with the fact
that satellite accretion is a generic feature of hierarchical models
of structure formation suggests that the model described in this study
should be applicable to at least some of the UFDs in the LG.

According to the impulsive approximation (which should be
valid in our simulations for the purposes of dimensional analysis),
the energy injected at each pericentric passage is given by $\Delta E
\propto M_{\rm host}^2 M R^2 V_{\rm rel}^{-2}$, where $M_{\rm host}$
is the mass of the host, $R$ is a characteristic radius of the dwarf
(related to the distance from the center of the dwarf where $V_{\rm
  max}$ occurs, $r_{\rm max}$), $M$ is a characteristic mass related
to the mass of the dwarf within $r_{\rm max}$, and $V_{\rm rel}$ is
the relative velocity between the two galaxies at the pericenter of
the orbit (Binney \& Tremaine 2008). Given that $M_{\rm host}$ is the
  same and $V_{\rm rel}$ is fairly similar for a given orbit, $\Delta
  E \propto M R^2$.

By virtue of the virial theorem, the energy content of the dwarf
should scale as $E \propto M^2/R$. Hence, the fractional increase in
energy caused by the tidal shocks is given by $\Delta E / E \propto
R^3 / M$. Increasing cusp slopes correspond to smaller
$R^3 / M$, and thus to smaller $\Delta E / E$. This explains why
for $\alpha=1$ the dwarfs experience a moderate amount of mass loss
and survive, while they are completely disrupted in the $\alpha=0.2$
case which is associated with the largest $\Delta E / E$. On the other
hand, the magnitude of tidal shocks when $\alpha = 0.6$ lies in between
the two aforementioned cases, explaining naturally the evolution of
these dwarfs towards UFDs.

Our simulations produce UFDs with $M_V > -5$ mag, $\mu_V =$ 28-29 mag arcsec$^{-2}$,
and $r_{1/2}=0.1$-$0.2$ kpc. These
observational properties are very similar to those of the
observed population of MW
UFDs.  This is valid for dwarfs evolved on fairly tight orbits, consistent with distances
interior to 100 kpc of 6 out of the 11 UFDs discussed in e.g. Kalirai
et al. (2010), whereas on less tight orbits remnants resembling
classic dSphs still result even for these shallow profiles.

One may wonder if the similarities extend to other properties of UFDs,
such as their inferred high DM content. The global
$M/L$ values at a scale of 1 kpc in both simulated UFDs are of the
order of 10-20 $M_\odot/L_\odot$. Admittedly, these $M/L$ ratios are lower than the values normally
quoted for UFDs (Kalirai et al. 2010). Note, however, that the
progenitors could have started out with most of the cold baryons in
the gaseous phase. Then ram pressure stripping aided by heating from
the ionizing background would remove most of them before they turned
into stars, possibly leaving behind an object with the same halo mass as
our UFD-like remnants but lower stellar mass.
These processes, in association with tidal stirring,
have already been shown to explain the most DM-dominated among classic
dSphs, such as Draco (Mayer et al. 2007).  In addition,
kinematic samples of UFDs are likely contaminated by tidally stripped
and MW stars, which could artificially enhance the measured $M/L$.

Other models for the origin of UFDs, such as the reionization
fossil scenario (Ricotti \& Gnedin 2005; Gnedin et al. 2008), would
not produce a correlation between UFDs and orbital distance,
but could explain the existence of UFDs at large distances
from the primaries, like Leo T (Irwin et al. 2007). In
such models the halo mass of the progenitor would be lower ($<10^8
M_{\odot}$). The different progenitor halo masses should result in
different overall numbers of the UFDs at $z=0$, both because halo mass
affects survival rate and the number of subhalos at different mass
scales are different.  The statistics of the population of UFDs
combined with orbital information could thus be a way to constrain
formation models.

\acknowledgments

This research was partially supported by the Polish National Science
Centre under grant NN203580940. S.K. is supported by the Center for
Cosmology and Astro-Particle Physics at The Ohio State University.
The numerical simulations were performed at
the Ohio Supercomputer Center (http://www.osc.edu).

\end{document}